\documentstyle[epsf]{l-aa}
\font\smallrm=cmr8
\def\lfir{$L_{\rm FIR}$}
\newcount\fignumber\fignumber=1
\def\nfig{\global\advance\fignumber by 1}
\def\fignam#1{\xdef#1{\the\fignumber}}
\fignam{\FMOL}          \nfig           
\fignam{\FVMR}          \nfig           
\fignam{\FHCNFIR}               \nfig           
\fignam{\FCOFIR}                \nfig           
\def\infig#1#2#3{\epsfxsize=#3cm \centering{\mbox{\epsfbox{#2}}}}

\newcount\tabnumber\tabnumber=1
\def\ntab{\global\advance\tabnumber by 1}
\def\tabnam#1{\xdef#1{\the\tabnumber}}
\tabnam{\TPARAM}        \ntab 
\tabnam{\TFLUX}         \ntab 
\tabnam{\TOPTLINE}      \ntab 
\tabnam{\TNH2}          \ntab 
\newcount\eqtnumber\eqtnumber=1
\def\eqtnum{{\the\eqtnumber}\global\advance\eqtnumber by 1}
\newcount\neqtnumber\neqtnumber=1
\def\neqt{\global\advance\neqtnumber by 1}
\def\eqtnam#1{\xdef#1{\the\neqtnumber}}
\begin{document}

\thesaurus{03(11.09.1 NGC~3049, NGC~5430, NGC~6764; 11.09.4; 11.11.1; 
11.16.1;
11.19.3)}
   \title{Starbursts in barred spiral galaxies}

   \subtitle{II. Molecular and optical study of three Wolf-Rayet galaxies
\thanks{Based on observations obtained at the 2 meter telescope of 
Observatoire 
du Pic du Midi, the 1.93 meter telescope of Observatoire de 
Haute-Provence,
both operated by INSU (CNRS), and at the 30 meter radiotelescope on Pico 
Veleta, operated by IRAM.}
	     }

   \author{T. Contini \inst{1} \and
	   H. Wozniak \inst{2} \and
	   S. Consid\`ere \inst{3} \and 
	   E. Davoust \inst{1} 
	   }

   \offprints{T. Contini, contini@wise.tau.ac.il}

   \institute{
Observatoire Midi-Pyr\'en\'ees, UMR 5572, 14 Avenue E. Belin, F-31400 
Toulouse,
France 
\and Observatoire de Marseille, URA CNRS 237, 2 Place Le Verrier, F-13248 
Marseille Cedex 4, France
\and Observatoire de Besan\c{c}on, EP CNRS 123, B.P. 1615, F-25010 
Besan\c{c}on Cedex, France
	     }

   \date{Received 27 November 1995; Accepted 22 January 1997}

   \maketitle

   \begin{abstract}
We have searched for dense molecular gas in three barred spiral galaxies with 
young starbursts, NGC~3049, 5430 and 6764, which are known Wolf-Rayet 
galaxies.  We detected HCN in the latter two, and CS was marginally detected 
in NGC~6764.  

The dense  molecular gas contents of the three galaxies are compared to those 
of other galaxies and to other indicators of star formation.  The HCN 
luminosities (relative to the CO and far infrared ones) in these galaxies with 
very young starbursts are consistent with those observed in galaxies with 
older starbursts and in normal galaxies, and so are our upper limits to the CS 
intensities (relative to CO).  

The starburst ages evaluated from our spectrophotometric observations are in 
the range 3.4 to 6.0 Myr.  A circum-nuclear ring is apparent on our images of 
NGC 5430, the galaxy with the oldest central starburst; this galaxy also has 
the widest molecular lines.  The central star formation rates derived from the 
H$\alpha$ luminosity are consistent with those expected from the global FIR 
luminosities, and are correlated with the HCN luminosities. 

Finally, an independent estimate of the H$_2$ column density is obtained by 
optical spectrophotometry; it leads to a H$_2$ column density to CO intensity 
ratio which is about 2 to 3 times lower than the standard value, because  
the CO intensities of the three galaxies are higher than average, relative to 
their far infrared fluxes. 
   

\end{abstract}

\section{Introduction}

\begin{table*}
\caption[~\TPARAM]{Basic data and global parameters of the three galaxies. D 
is the distance, $B_{\rm tc}$ is the apparent corrected blue magnitude and $i$ 
is the inclination.  The logarithms of total HI mass, and of far infrared and 
blue total luminosities are given in the last three columns} 

\begin{flushleft}
\begin{tabular}{clccccccrr}
\noalign{\smallskip}
\hline
\noalign{\smallskip}
Galaxy&Type&\multicolumn{2}{c}{Coordinates (J2000)}&D&$B_{\rm 
tc}$&$i$&M(HI)
&$L_{\rm FIR}$&$L_{\rm B}$ \\
&&$\alpha$&$\delta$&(Mpc)&(mag)&(\degr)&log(M$_{\sun}$)&log(L$_{\sun}$)
&log(L$_{\sun}$) \\
\noalign{\smallskip}
\hline
\noalign{\smallskip}
NGC 3049& SBb &$09^{\rm h}54^{\rm m}49\fs7$& +09\degr16\arcmin19\arcsec&19.6& 12.97& 50.6& 
9.10&  9.31& 9.57 \\
NGC 5430& SBb &$14^{\rm h}00^{\rm m}45\fs7$& +59\degr19\arcmin42\arcsec&42.7& 12.43& 51.7& 
9.58& 
10.63& 10.46 \\
NGC 6764& SBbc&$19^{\rm h}08^{\rm m}16\fs7$& +50\degr55\arcmin54\arcsec&35.6& 11.89& 50.1& 
9.65& 
10.23& 10.52 \\
\noalign{\smallskip}
\hline
\end{tabular}
\end{flushleft}

\end{table*} 
 
Despite numerous studies of starburst galaxies, many questions remain about 
the starburst phenomenon, its triggering and evolution.  Gravitational 
interactions and mergers seem to play a major role in triggering starbursts, 
but violently interacting galaxies are not necessarily the seat of starbursts 
(Bushouse 1986) and most starbursts seem to be isolated (Contini 1996, Coziol 
et al. 1996b). Numerical simulations have shown that the bar plays a major 
role in this process, by efficiently funneling molecular clouds toward the 
inner few kiloparsecs of galaxies (Nogushi 1988, Friedli \& Benz 1993). This 
is confirmed by radio continuum observations (e.g. Puxley et al. 1988). 
However, the link between bar and far infrared luminosity (which traces young 
massive stars) in starburst galaxies remains controversial (Hawarden et al. 
1996). 

Molecular clouds obviously play a crucial role in the process of star 
formation, and their properties have been extensively studied, via millimeter 
observations of the CO molecule.  A strong far infrared (FIR) luminosity has 
generally been a successful criterion for detection of the molecule in 
external galaxies, confirming the link between the two indicators of star 
formation.  A CO--$N$(H$_2$) conversion factor has been proposed (Strong et 
al. 1988) and validated by observations of normal as well as starburst 
galaxies (e.g. Sage et al. 1990); further studies have shown that it depends 
on metallicity (Wilson 1995, Arimoto et al. 1996).  But its validity has 
recently been questioned (e.g. Nakai \& Kuno 1995). 

It has recently been emphasized that CO only traces low-density molecular gas 
and several searches for dense (n(H$_{2}$)~$>10^4$~cm$^{-3}$) molecular 
clouds in normal and starburst galaxies, generally selected to be strong CO 
emitters, via detection of HCN, CS, HCO$^+$, and other molecules, have been 
initiated (e.g. Mauersberger et al. 1989, Nguyen-Q-Rieu et al. 1992, Helfer 
\& Blitz 1993, Aalto et al. 1995).  It turns out that bulges of normal as well 
as starburst galaxies contain large quantities of dense gas, and that a 
threshold in the surface density of dense gas does not seem to be required for 
violent star formation (Helfer \& Blitz 1993). 

In view of these mixed results, we have adopted a global view on starbursts in 
galaxies, based on multi-wavelength observations of a large and homogeneous 
sample of barred starburst galaxies.  Such an approach should enable us to 
establish quantitative relationships between the properties of starbursts 
(age, star formation rate and initial mass function), the neutral (atomic and 
molecular) gas content and the morphology of the host galaxies.  It has in 
fact already given rise to new and original results (Contini et al. 1995, 
Contini 1996, Contini et al. 1996a, Coziol et al. 1996a). 

In this paper, we combine millimetric observations of the dense molecular gas 
with optical images and long-slit spectroscopy of a few examples of very young 
starbursts galaxies, namely Wolf-Rayet galaxies, in order to investigate the 
properties of the dense gas of these galaxies in relation to their optical 
properties.  The images provide morphological and photometric informations on 
the central regions and the bar, and the spectra are used to determine the 
starburst ages and star formation rates as well as an independent estimate of 
the column density of H$_2$. A preliminary report on this research project has 
been published by Contini et al. (1996b). 

\section{The starburst phenomenon in Wolf-Rayet barred galaxies} 

Wolf-Rayet galaxies are characterized by the presence of a large (10$^2$ to 
10$^3$) number of Wolf-Rayet stars, which can only be explained by a very 
recent starburst episode, between 3 and 6 Myr old (Vacca \& Conti 1992, Maeder 
\& Meynet 1994), and probably with a nearly flat initial mass function 
(Contini et al. 1995).  A large number of very massive ($M > 30 M_{\odot}$) OB 
stars, which are the progenitors of Wolf-Rayet stars, are likely to form under 
rather unusual circumstances; one might thus expect the interstellar medium of 
this type of galaxies to be different from those of galaxies with older and/or 
more moderate starbursts. 

We selected our sample from the catalogue of Conti (1991). In order to 
optimize detection of the molecular lines, we chose three barred galaxies with 
a low redshift and a large far infrared flux, NGC~3049, 5430 and 6764. 
Fundamental parameters for the three galaxies, derived from the Lyon Meudon 
Extragalactic database (LEDA), are presented in Table~\TPARAM. 

Wolf-Rayet galaxies do not form a homogeneous class of galaxies, and span a 
wide range in luminosity, size and morphological type; in this respect, the 
three selected galaxies do show differences. NGC~5430 and 6764 are giant 
galaxies, whereas NGC~3049 is of low luminosity. The starburst of NGC~6764 
occurs in the center of the galaxy, but the Wolf-Rayet regions of NGC~3049 and 
5430 are extranuclear, the former is 2.5\arcsec\ South-West of the  nucleus 
along the bar, the latter at the South-Eastern end of the bar (12\arcsec\ East 
and 18\arcsec\ South of the nucleus). The two latter galaxies are also 
Markarian galaxies (Mkn 710 and 799 respectively). One property shared by all 
three galaxies (and by most known Wolf-Rayet barred spiral galaxies; Contini 
et al. 1995) is their high inclination. 

\begin{table*}
\caption[~\TFLUX]{Heliocentric radial velocities (V), line widths (FWHM)
and integrated intensities ($I$) of the molecular lines and HI data} 
{\small

\begin{flushleft}
\begin{tabular}{lcrrcrcccc}
\noalign{\smallskip}
\hline
\noalign{\smallskip}
Galaxy  &\multicolumn{3}{c}{CO(2 $\rightarrow$ 1)}&\multicolumn{3}{c}
{HCN(1 $\rightarrow$ 0)}&CS(3 $\rightarrow$ 2)&\multicolumn{2}{c}{HI}\\ 
&V&FWHM&$I_{\rm CO}$&V&FWHM&$I_{\rm HCN}$&$I_{\rm CS}$&V&FWHM\\
&(km s$^{-1}$)&(km s$^{-1}$)&(K km s$^{-1}$)&(km s$^{-1}$)&(km 
s$^{-1}$)&(K km
s$^{-1}$)&(K km s$^{-1}$)&(km s$^{-1}$)&(km s$^{-1}$)\\
\noalign{\smallskip}
\hline
NGC 3049&1500$\pm$1&73$\pm$3&9.2$\pm$0.3&&&$<$0.3&$<$3.2&1494&199\\
\noalign{\medskip}
NGC 
5430&2981$\pm$1&389$\pm$3&81.6$\pm$0.5&2985$\pm$17&248$\pm$29&1.4$\pm$0.2
&$<$2.5& 2965&313\\
(NW)    & 2893$\pm$1& 141$\pm$2& 38.4$\pm$0.5&&&&     &     \\
(SE)    & 3069$\pm$1& 131$\pm$2& 43.2$\pm$0.5&&&&     &     \\
WR region& 3074$\pm$3&  97$\pm$5&  6.1$\pm$0.5&&&$<$0.5&$<$0.9&    &     
\\
\noalign{\medskip}
NGC 6764& 2431$\pm$1& 148$\pm$2& 25.6$\pm$0.5&2422$\pm$11& 139$\pm$25& 
0.8$\pm$0.2&(1.0)& 2416&276\\
\noalign{\smallskip}
\hline
\end{tabular}
\end{flushleft}

}
\end{table*}

\section{Observations and data reduction}

The radio observations were obtained at the IRAM 30 meter radiotelescope on 
November 27 and 28, 1994. We observed in single sideband at three frequencies 
simultaneously, CO(2$\rightarrow$1) at 230.5 GHz with the 1.3mm (230G1) SIS 
receiver, CS(3$\rightarrow$2) at 147.0 GHz with the 2mm receiver, and 
HCN(1$\rightarrow$0) at 88.6 GHz with the 3mm SIS receiver. Upper sideband 
rejection was 10 to 15, 7 to 9 and 30 dB for CO, CS and HCN respectively. As 
backends, we used an autocorrelator for CO and filter banks of 512 1-MHz 
channels for the two other lines. The beamwidth of the 30 meter antenna is 27, 
16 and 11\arcsec\, and the main beam efficiency 0.75, 0.52 and 0.39 in order 
of increasing frequency (Kramer \& Wild 1994). The weather was remarkably good 
during the run, with average zenith opacities of 0.08 at 88 and 147 GHz and 
0.17 at 230 GHz. The average system temperatures were 220, 320 and 400 K in 
order of increasing frequency. The observations were made in wobbler mode with 
a throw of 2 to 4 arcmin. Pointing was done on quasars every 2 or 3 hours and 
found to be stable; focus was done on planets near sunrise and sunset. On-off 
integration times were about 3.5 hours for NGC~3049 and the Wolf-Rayet region 
of NGC~5430, and over 5 hours for the centers of NGC~5430 and 6764. 

We checked the calibration of the CO brightness temperatures of NGC~5430 and 
6764 by observing the source IRC+10216. For NGC~3049, we observed W51E1, but 
the predicted lines were not seen (presumably because the catalogued lines 
arose in the USB at the time the catalogue was done), and the calibration 
could not be tested; the brightness temperatures and derived integrated 
intensities are thus uncertain by about 30\%. It was not necessary to 
calibrate the HCN brightness temperatures, since USB rejection was good at the 
corresponding frequency. The observations were reduced with CLASS. 

CCD images of the three galaxies were obtained during two runs at the 2 meter 
telescope of Observatoire du Pic du Midi, with a 1000$\times$1000 Thomson CCD 
(pixel size 0.20\arcsec\ on the sky). Images of NGC~3049 and 5430 were 
obtained in January of 1992, with respective exposure times 20 and 30 min. in 
V and 15 and 20 min. in R. Images of NGC~6764 were obtained in August of 1992, 
with exposure times 40 min. in B and 20 min. in V. 

The calibration of the zero point of the magnitude scale was done by indirect 
procedures. The photometric constants in V were obtained by the measure of 
stars from the Guide star Catalogue (GSC) or using published aperture 
photometry (Longo \& de~Vaucouleurs 1983). For NGC~5430, we fitted our R 
surface brightness profile to the r-Gunn profile of Kent (1985), taking into 
account the color equations between Johnson and Gunn's photometric systems. We 
also verified that the total magnitude of each galaxy agreed reasonably well 
with published values.  The accuracy of the zero point is thus 0.1~mag. 

The spectroscopic observations were obtained in January 1992 and June 1996 at 
the 1.93 meter telescope of Observatoire de Haute-Provence, with the Carelec 
spectrograph (Lema\^\i tre et al. 1990), at a spectral resolution of 
260\AA/mm.  During the first run, we took one 30 min. spectrum of NGC 3049 
along the major axis. The slit width was 2.2\arcsec. For flux calibration, we 
observed the standard stars GD~140 (Massey et al. 1988) and BD~ +26~2606 (Oke 
\& Gunn 1983). In the second run, we took six 25 min. spectra of NGC 5430 and 
2 of NGC 6764. The slit width was 2.8\arcsec, and the standard star for flux 
calibration was HD 192281. The spectra were reduced with MIDAS. 

\section{Results}

\subsection{Molecular line profiles}

The molecular line profiles are presented in Fig.~\FMOL.  HCN was detected in 
the centers of NGC~5430 and 6764, but not in the Wolf-Rayet region of 
NGC~5430, nor in NGC~3049. CS was marginally detected in the center of 
NGC~6764 only.  The heliocentric radial velocities, line widths and integrated 
intensities in each line of the three galaxies were estimated by fitting 
gaussians or saddle-shapes to the profiles; they are summarized in 
Table~\TFLUX, where the HI velocities and linewidths are also given for 
comparison.  The integrated intensity in each line is $I = \int{T_{\rm 
mb}dv}$, where $T_{\rm mb}$ is the main-beam temperature.  The quoted 
uncertainties on the integrated intensities are internal and do not take into 
account pointing and calibration inaccuracies. For undetected lines, we give a 
2$\sigma$ upper limit of the integrated intensity. The integrated intensities 
must be multiplied by 4.89, 4.73 and 4.77 (at 88, 147 and 230 GHz 
respectively) in order to be converted to Jy km s$^{-1}$. Since NGC~5430 and 
6764 are fairly distant compared to most galaxies where HCN has been detected, 
the HCN integrated intensities given in this paper are among the lowest ever 
reported. 

\begin{figure*}
\infig{21}{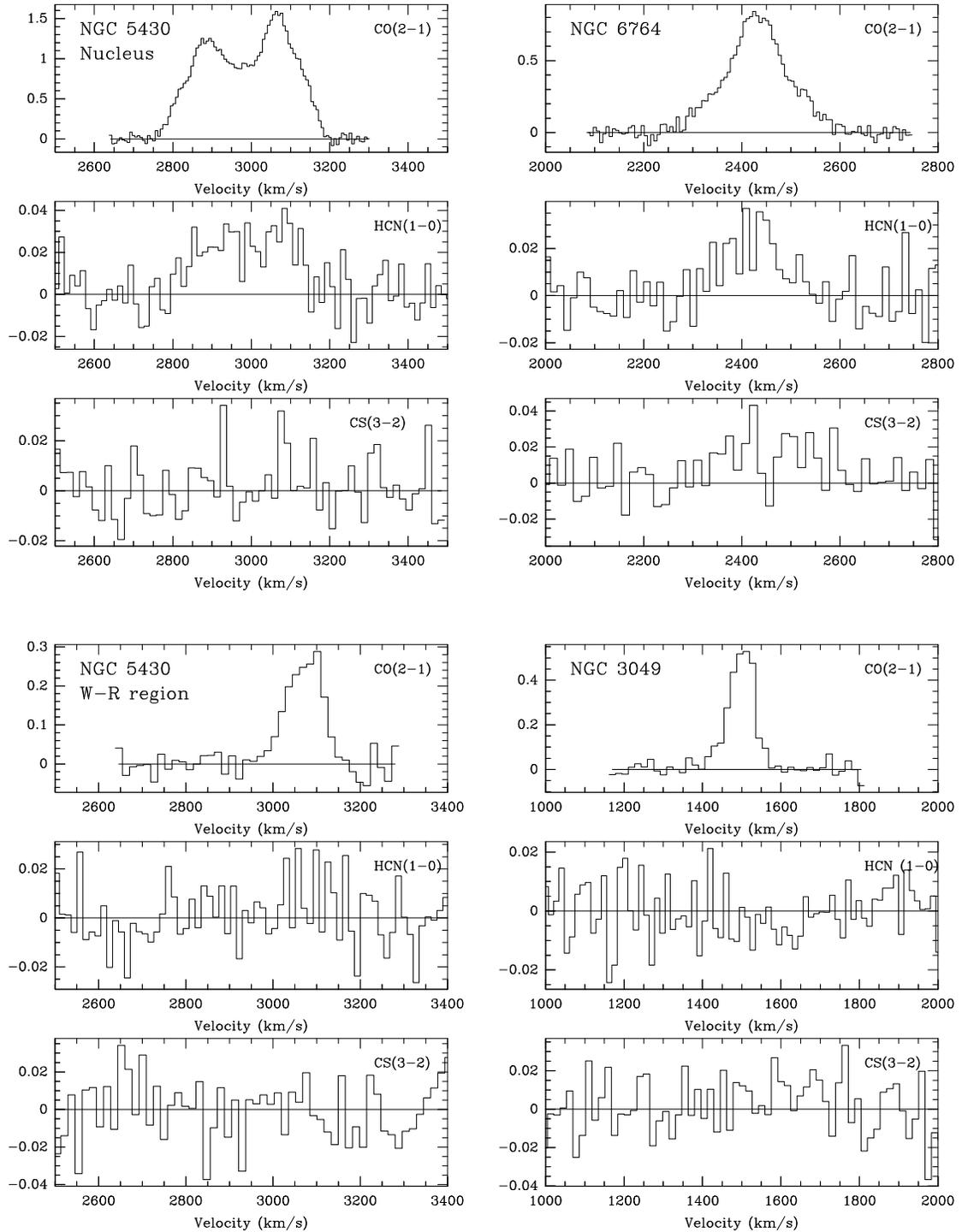}{16}
\caption[\FMOL]{Molecular line profiles (in Jy) of the three lines in the 
three galaxies. Top left : center of NGC 5430, bottom left : WR region of NGC 
5430, top right : NGC 6764, bottom right : NGC 3049.  For each galaxy, we 
present from top to bottom the CO(2$\rightarrow$1), HCN(1$\rightarrow$0) and 
CS(3$\rightarrow$2) profiles}   
\end{figure*} 

\begin{figure*}
\infig{21}{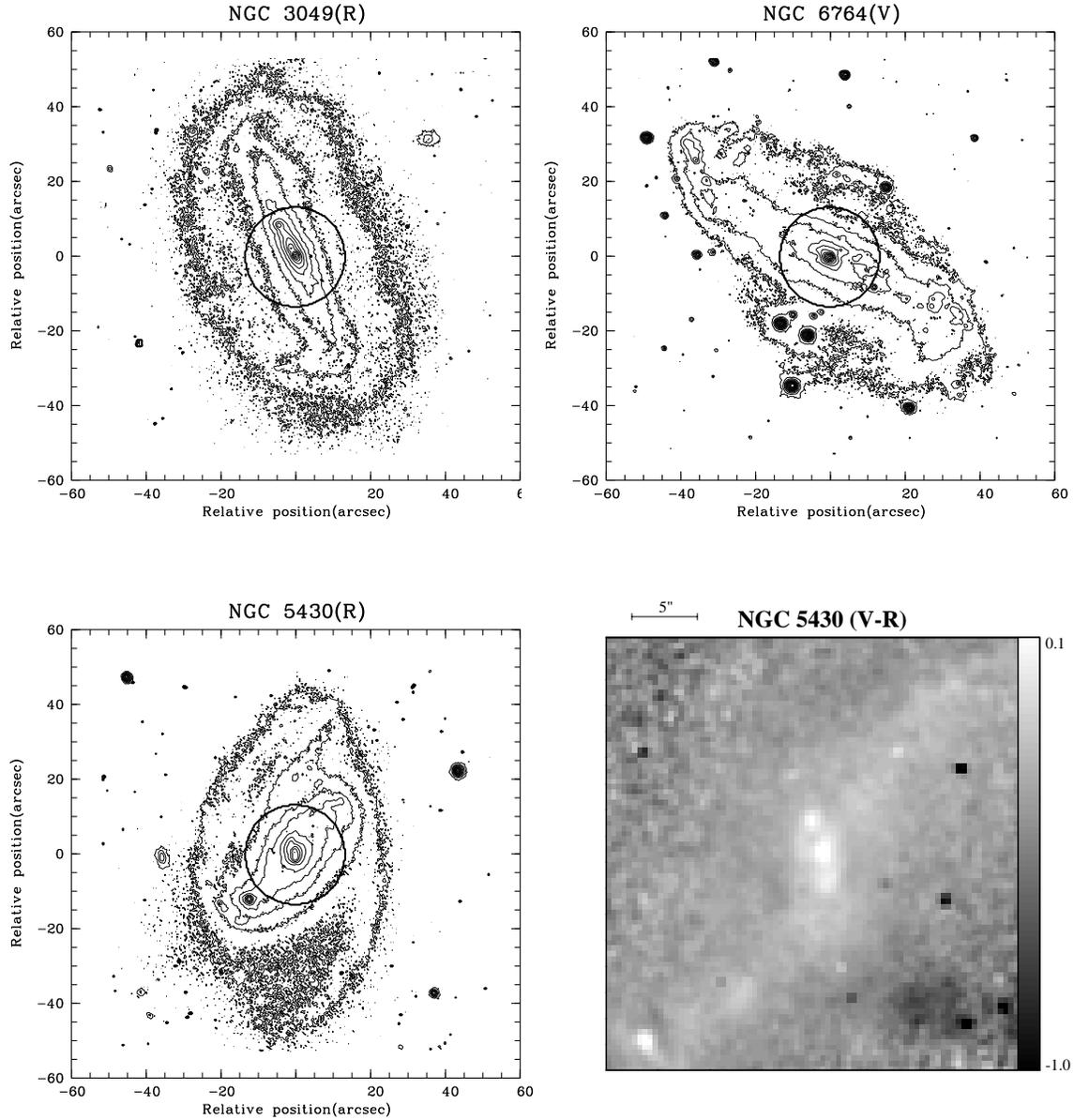}{16}
\caption[\FVMR]{Isophote maps for the R band of NGC~3049 and NGC~5430 and the 
V band of NGC~6764.  The spacing of the isocontours is 0.5~mag. The faintest 
isophote levels are 22, 21 and 22.5 mag arcsec$^{-2}$ respectively.  The size 
of the HPBW at 87 MHz is indicated by a circle. The lower right panel shows 
the  V--R color map of the innermost region of NGC~5430. On this  map, the 
lower left bright spot is the Wolf-Rayet region. North is up and East  to the 
left} 
\end{figure*} 

Our detection of CO in NGC~3049 is the first one reported for this galaxy. The 
CO line profile of NGC~3049 is in fact not perfectly gaussian; it shows an 
excess of gas at higher velocities.  It is presumably due to the bright HII 
region centered 8\arcsec\ to the North-East. 

The central CO(2$\rightarrow$1) profile of NGC~5430 has previously been 
observed by Kr\"ugel et al. (1990), with the same instrument; the profile and 
integrated intensity are in good agreement with ours. They also observed the 
CO(1$\rightarrow$0) and $^{13}$CO(1$\rightarrow$0) lines in the center of this 
galaxy. In the center of NGC~5430, the CO line profile is double peaked. This 
is the typical signature of a gaseous disk or ring.  Such a structure is very 
likely associated with the nuclear hot-spots seen in our V--R color map 
(Fig.~\FVMR). 

In the Wolf-Rayet region of NGC~5430, CO has a mean velocity of 
3074 km s$^{-1}$, which means that the South-East is receding, and that the 
near side of the galaxy is the Eastern side.  The fact that this region, which 
is 19.3\arcsec\ from the center, has the same mean velocity ($\simeq$ 3070 km 
s$^{-1}$) as the high velocity peak of the central profile indicates that 
most of the bar is in the differentially rotating part of the galaxy. 

The high-velocity peak of CO is larger than the low-velocity one. The HI 
profile (Roth et al. 1991) of this galaxy is also asymmetric, but the highest 
peak is at the other end of the profile; there appears to be less neutral 
hydrogen where there is more CO, on the SE side of the galaxy, where the HII 
region with Wolf-Rayet emission lies. Another surprising related fact is that 
the FWHM of the CO line is {\it larger} than that of the HI line (389 vs 313 
km s$^{-1}$), because high velocity HI is definitely lacking. 

The central CO(2$\rightarrow$1) profile of NGC~6764 has been published by 
Eckart et al. (1991), also using the IRAM 30 meter antenna; they find a rather 
high integrated intensity of 35 K km s$^{-1}$ compared to ours, perhaps
because they used a slightly different position. 
The CO profile of this galaxy is similar to that of NGC~3049, and indicates 
that the molecular gas is in solid-body rotation in the central 11\arcsec\ of 
this galaxy as well. CO(1$\rightarrow$0) also appears to have a constant 
velocity gradient across the bar (Eckart et al. 1991), in agreement with the 
results of Rubin et al. (1975) for the ionized gas \footnote{in that paper, 
East and West must be interchanged (Rubin, private communication).}.  

\subsection{Optical properties} 

The geometrical and photometric properties of the three galaxies were obtained 
by fitting ellipses to the isophotes of the CCD images in the two bands, using 
a method described in Wozniak et al. (1995), and analyzing the resulting 
surface brightness and ellipticity profiles. 

NGC~3049 has a very thin bar ($b/a \sim 0.18$), 
of constant surface brightness, and a small bulge. The apparently elongated 
nucleus ($r<10$\arcsec) shows isophotes which are twisted by the HII region 
close to the nucleus. The separation between the nucleus and this spot is 
2.5\arcsec. There is a bright spot of star formation in the North-East 
(labeled ``b'' by Mazzarella  \& Boroson 1993).  An inner ring surrounds the 
bar. 

NGC~5430 displays an elongated nucleus. This shape is due to at least two 
bright spots on a circumnuclear ring.  This ring is clearly visible on the 
V--R color map (Fig.~\FVMR). Its semi-major axis is 3.5\arcsec. The V--R 
colors of the nucleus and the spots are respectively $\sim 0.2$ and $\sim 0$ 
to $-0.1$ while in the bar outside the dust lanes it is $\sim -0.3$. The 
Wolf-Rayet region has almost the same color as the nucleus (V--R $\sim 0.25$).  
The bar does not have a constant surface brightness, contrary to those of the 
two other galaxies. There is also a 3-armed spiral structure, the southern arm 
being more elongated than the opposite one. 

NGC~6764 has a thin bar of roughly constant surface brightness and a small 
bulge, like NGC~3049. The ellipticity of the bar increases outward to a 
maximum of 0.8, but its position angle remains constant. 

\subsection{Star formation rate and age of the starbursts} 

We used long-slit spectra to estimate the spectrophometric properties of six 
starburst regions located in our three Wolf-Rayet galaxies. For each starburst 
we measured the reddening coefficient $C_{{\rm H}\beta}$, the absolute 
dereddened H$\alpha$ intensity $I$(H$\alpha$), the H$\beta$ equivalent width 
W(H$\beta$), and the oxygen abundance, O/H, which is a metallicity indicator. 
The procedure for analyzing our low-dispersion spectra is described in Contini 
(1996) and Contini et al. (1995). The results are given in Table~\TOPTLINE. 

The H$\alpha$ luminosity $L$(H$\alpha$) can be used to estimate the 
star formation rate (Kennicutt 1983), because it is directly proportional to 
the number of ionizing photons produced by massive and hot OB type stars 
(Osterbrock 1989).  The predicted central star formation rates are 0.7, 8.4 
and 3.7 M$_{\odot}$ yr$^{-1}$ for NGC~3049 (including the NE and SW knots), 
NGC~5430 and NGC~6764 respectively, to be compared with the mean value of 4 
M$_{\odot}$ yr$^{-1}$ for the {\it total} star formation rate in normal Sb-Sbc 
galaxies (Kennicutt 1983). 

In starburst galaxies, the integrated FIR luminosity \lfir\ is mostly due to 
dust heated by UV radiations from young stars, rather than by absorption of 
visible radiation from older stars in the disk. It can thus also be 
used to estimate the star formation rate, and good correlations have been 
found between $L$(H$\alpha$) and \lfir\ for samples of IRAS galaxies (Devereux 
\& Young 1991, Sauvage \& Thuan 1992, Coziol 1996).  In a sample of 79 barred 
starburst galaxies, Contini (1996) found a nearly linear relation between  
$L$(H$\alpha$) in the center of the galaxies and their global \lfir\, with 
$L$(H$\alpha$) $\propto$ \lfir$^{0.92}$.  Our three Wolf-Rayet galaxies follow 
this relation quite closely, given the uncertainties on the absolute 
intensities.  This means that, although the central star formation rates of 
our three galaxies are high compared to those in normal spirals, they are 
average compared to those of other barred starburst galaxies. 

There has been some confusion in the literature about the nature of the 
nuclear activity of NGC~6764 (see Contini et al. 1996b for the various 
classifications).  Our new spectra clearly show that the center as well as the 
circum-nuclear region are the seat of LINER activity. 

We estimated the ages of the starbursts by comparing our spectrophotometric 
data with predictions of the evolutionary synthesis model of Cervi\~no \& 
Mas-Hesse (1994). A good accuracy ($\pm$ 0.5 Myr) is obtained by 
using W(H$\beta$), for a given value of O/H.  

In NGC 3049, the starburst age increases from 3.4 Myr in the Wolf-Rayet region 
(SW) to 5.6 Myr in the most distant (NE) of the bright HII regions along the 
bar. The metallicity is roughly constant along the bar, and the internal 
extinction is more important along the bar than in the nucleus, presumably 
because of dust.  NGC 5430 also experienced two starbursts in the recent past, 
one in the Wolf-Rayet region at the end of the bar (SE), 4 Myr ago, and one 
in the nucleus (C), about 6 Myr ago (with a larger uncertainty of about 1 Myr, 
due to the uncertainty on the metallicity).  Finally, the age of the 
starburst in the nucleus of NGC~6764 is 5.2 Myr.  The derived ages in the 
starbursts of the three Wolf-Rayet galaxies are listed in Table~\TOPTLINE. 

\begin{table}
\caption[~\TOPTLINE]{Spectrophotometric data and derived starburst ages in the 
3 galaxies. $I$(H$\alpha$) is in 10$^{-14}$ erg s$^{-1}$ cm$^{-2}$}. 
\smallrm{ 

\begin{flushleft}
\begin{tabular}{clrrrccc}
\noalign{\smallskip}
\hline
\noalign{\smallskip}
NGC&\multicolumn{2}{c}{Region}
&$I$(H$\alpha$)
&W(H$\beta$)
&$C_{{\rm H}\beta}$
&O/H
&age\\
&&(\arcsec)
&
&(\AA)
&(mag)
&($\sun$)
&(Myr)\\
\noalign{\smallskip}
\hline
\noalign{\smallskip}
3049&NE& 8.0& 12&--16&0.74&1.7&5.6\\
    &C & 0.0& 20&--29&0.32&1.2&4.7\\
    &SW& 2.5& 30&--38&0.23&1.3&3.4\\
\noalign{\medskip}          
5430&C & 0.0&103& --5&1.63&1.3&6.0\\
    &SE&22.0& 37&--41&0.56&1.2&4.0\\ 
\noalign{\medskip}
6764&C & 0.0& 75&--21&0.87&0.8&5.2\\
\noalign{\smallskip}
\hline
\end{tabular}
\end{flushleft}

} 
\end{table} 

\section{Dense molecular gas contents in Wolf-Rayet galaxies} 

We detected HCN(1$\rightarrow$0) in the center of NGC~5430 and of NGC~6764, 
but not in NGC 3049, and CS(3$\rightarrow$2), although marginally, in NGC 6764 
only.  In order to evaluate quantitatively whether these galaxies, which have 
unusually young starbursts, are also unusual in their dense molecular gas 
contents, we compare our measurements with those published for other surveys 
of HCN and CS in external galaxies.   

In the following discussion, we make the simplifying assumption that the 
luminosity of these lines is a measure of the mass of dense (n(H$_{2}$)~$> 
10^4$~cm$^{-3}$)  gas.  Although this assumption is debatable (Aalto et al. 
1995), it is generally adopted in most studies of this kind, implicitly, or 
explicitly by a model (e.g. Solomon et al. 1990). 

\subsection{$L_{\rm HCN}$ vs \lfir}

We first compare the HCN luminosity of the galaxies with their total far 
infrared luminosity, \lfir. 
In 
other words, we compare the mass of dense gas from which stars are born to the 
number of recently formed stars. 

To this end, we use the HCN surveys of Nguyen-Q-Rieu et al. (1992), Helfer \&  
Blitz (1993), and Solomon et al. (1992), and HCN observations of individual  
galaxies, NGC~4945 by Henkel et al. (1994), and NGC 3256 by Casoli et al.  
(1992).  The comparison with our results is given in  Fig.~\FHCNFIR, which 
shows the dependence of the HCN luminosity  on total far-infrared luminosity.   
The comparison samples are divided into two categories, high resolution  
(26\arcsec\ to 28\arcsec\ HPBW) and low resolution (56\arcsec\ to 63\arcsec\  
HPBW) observations. The extranuclear Wolf-Rayet region of NGC 5430 is not 
plotted because the comparison samples only concern central regions of 
galaxies. 

\begin{figure}
\infig{12}{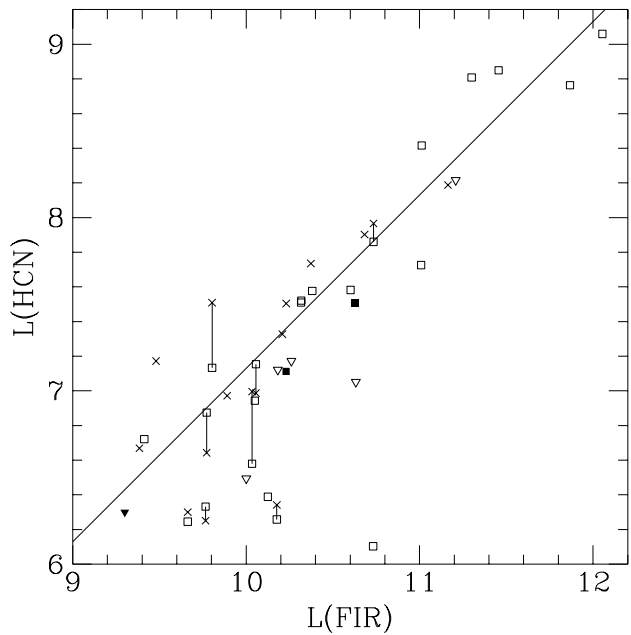}{8.8}
\caption[\FHCNFIR]{HCN luminosity (in K km s$^{-1}$ pc$^2$) vs far-infrared 
luminosity (in L$_{\sun}$).  Filled squares: NGC 5430 and 6764; filled 
triangle: NGC 3049 (upper limit in HCN); open squares and crosses: high 
resolution (26\arcsec\ HPBW) and low resolution (56 to 63\arcsec\ HPBW) data 
from the literature; open triangles: upper limits (in HCN luminosity) from 
the literature.  Data for the same galaxy observed in both modes are linked by 
a vertical line} 
\end{figure} 

The diagonal line is a linear fit (with slope unity) to the data in the region 
where there indeed seems to be a correlation ($L_{\rm FIR} > 10^{10.2}$ 
L$_{\sun}$). As no marked differences between the two comparison samples 
appear on this diagram, at least above the quoted limit, we assume that all 
the HCN emission is unresolved, even by the smaller beam.  All the nearby 
galaxies (NGC 253, M33, M82), are below this limit, and some of them are 
probably resolved.  

This correlation should be viewed with caution, as it might reveal more about 
the way the sample was selected (and about the Malmquist bias) than about 
physical properties of the galaxies. It can nevertheless be used to compare  
the {\it relative} behavior of the different galaxies. Our two detected  
galaxies are just below the diagonal line, and so is the upper limit derived  
for NGC 3049.  This means that {\it the HCN luminosities of our galaxies with 
young  starbursts are not unusually high}, and that NGC 3049 was not detected 
because  it is probably just below the detection capabilities of the 
instrument; the noise on the HCN spectrum of NGC 3049 in Fig.~\FMOL\ is at a 
level of  $T_{\rm mb}$ = 2mK. 

\subsection{$I_{\rm HCN}$ vs $I_{\rm CO}$}

\begin{figure}
\infig{12}{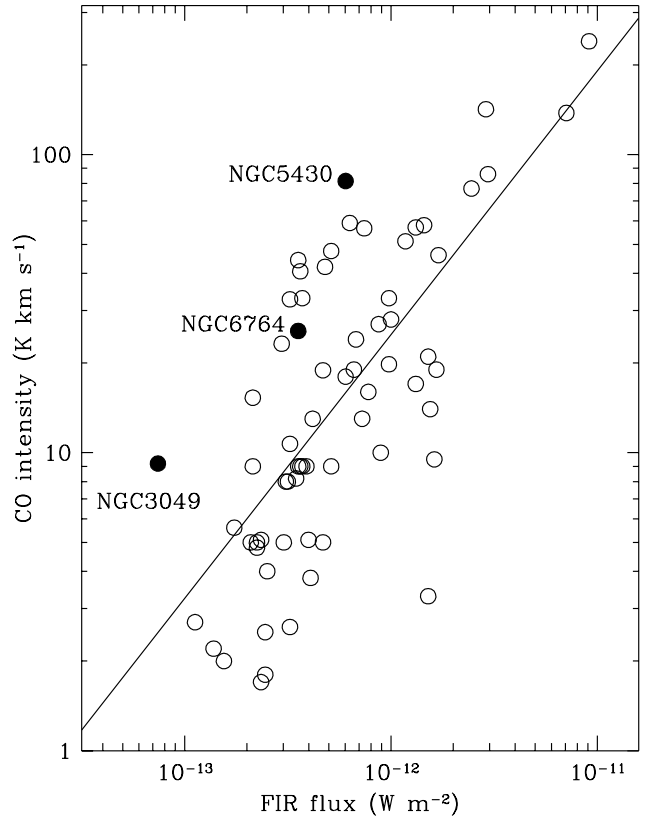}{8.8}
\caption[\FCOFIR]{CO intensity (in K km s$^{-1}$) vs far-infrared flux (in W 
m$^{-2}$).  Filled circles: NGC 3049, 5430 and 6764; open circles : 
comparison samples of Braine \& Combes (1992) and Contini et al. (1996a). 
The diagonal line is a least-squares fit to the data of the comparison samples} 
\end{figure} 

Another way of testing whether the dense molecular gas content of our galaxies 
is unusual is to measure the intensity ratio of HCN to CO, which provides an 
estimate of the molecular gas density.   We compare the values of this ratio 
in our galaxies with those in the sample of galaxies with unremarkable nuclear 
properties of Helfer \& Blitz (1993).  We cannot use our CO(2$\rightarrow$1) 
observations to this end, since the corresponding beam is much smaller than 
that of HCN. Instead, we use the CO(1$\rightarrow$0) observations of Kr\"ugel 
et al. (1990) for NGC~5430 ($I_{\rm CO}$ = 49.8 K~km~s$^{-1}$) and of Eckart 
et al. (1991) for NGC~6764 ($I_{\rm CO}$ = 25 K~km~s$^{-1}$).  The latter 
should be taken with caution, since we disagree with their measure of 
CO(2$\rightarrow$1) in the same galaxy. 

The ratios $I_{\rm HCN}$/$I_{\rm CO}$ are 0.028 and 0.032 for NGC~5430 and 
NGC~6764 respectively.  This is significantly (at least 3$\sigma$) below the 
mean value of 0.078 for other IRAM observations of bulges of galaxies (see 
Table~3 of Helfer \& Blitz 1993).  These ratios are in fact much closer to the 
value of 0.026 found in the disk of our own Galaxy (Helfer \& Blitz 1996)
or that of 0.03 to 0.06 in the disk of M51 (Kohno et al. 1996). 

However, Helfer \& Blitz (1993) noticed that the HCN--CO line ratio depends on 
the beam size, and is smaller for a larger beam. One could thus raise the 
objection that the HCN--CO ratio for a given beamwidth is distance dependent, 
if HCN is probably more centrally concentrated than CO, and that our ratios 
are underestimated, since our galaxies are fairly distant. 

But the velocity widths of HCN and CO are the same in NGC 6764, indicating a 
comparable spatial extent of both molecules.  HCN should thus perhaps be 
viewed as distributed in a dense core {\it and} in a diffuse disk; this 
is the case for M51 (Kohno et al. 1996), NGC 1068 (Helfer \& Blitz 1995)
and NGC 6946 (Helfer \& Blitz 1996).  Furthermore, our beam (27\arcsec) is 
rather small compared to that 
of smaller antennas (56 to 63\arcsec) used in some HCN surveys, and the 
distance effect, if any, should appear at larger distances than those of our 
galaxies.  We thus do not think that our two line ratios are 
underestimated, and we confirm that the HCN to CO line intensity ratios of 
the two galaxies are rather low compared to those of other 
galaxies observed with the same instrument.  

The non detection of HCN in the Wolf-Rayet region of NGC 5430 can be explained 
if, disregarding the different beamwidths for the two molecules, one assumes 
that the HCN intensity ratio between the center and the Wolf-Rayet region is 
the same as the CO intensity ratio. For the latter, we found a value of 13. 

\subsection{$I_{\rm CO}$ vs FIR flux}

We found above that the HCN to CO intensity ratio of our galaxies is well 
below average.  This suggests that their CO intensity must be unsually high, 
since their HCN luminosity is normal.  To check this possibility, we compare 
the CO intensities (relative to the FIR emission) of our galaxies with those of 
two other samples.  The CO(1$\rightarrow$0) integrated intensities of our 
galaxies are plotted vs their FIR flux on Fig~\FCOFIR, together with those of 
52 galaxies of Braine \& Combes (1992; see their Fig.~5a) and of 24 young 
starburst galaxies observed by Contini et al.~(1996a). On this figure, our 
three galaxies occupy the upper boundary; they indeed have a higher 
CO(1$\rightarrow$0) intensity than average (by a factor 2.5 to 4) for their 
FIR flux.  We have verified that the same is true for the CO(2$\rightarrow$1) 
intensities.  

\subsection{$I_{\rm CS}$ vs $I_{\rm HCN}$ and $I_{\rm CO}$}

We now turn to the contents of CS in our galaxies.  Helfer \& Blitz (1993) 
find a ratio of 2.4 between the integrated intensities of HCN(1$\rightarrow$0) 
and CS(2$\rightarrow$1) in their sample.  A ratio of that order is consistent 
with the marginal detection in NGC~6764, and with the non detection in 
NGC~5430. Intensity ratios of CS(2$\rightarrow$1) to CO(1$\rightarrow$0) of 
0.03--0.06 have been quoted in the literature (Sage et al. 1990, Helfer \& 
Blitz 1993).  Using once more CO(1$\rightarrow$0) intensities from the 
literature, and assuming that CS(3$\rightarrow$2)/CS(2$\rightarrow$1) is close 
to unity, we again find that our results are consistent with expectations.  
One should keep in mind that these are order of magnitude estimates only, 
since our detection of CS in NGC 6764 is marginal. 

\section{The column density of H$_2$ and the CO--$N$(H$_2$) conversion 
factor}                                      

We take advantage of the fact that molecular and optical spectroscopy are 
available for our galaxies to compare two methods for estimating the 
CO--$N$(H$_2$) conversion factor, a quantity of major importance for 
estimating the amount of hydrogen in molecular form in galaxies. 

In the extinction method, one takes the ratio of two quantities directly 
related to the galaxy.  The average column density of H$_2$ between us and the 
center of a galaxy with metallicity $Z$ (in units of $Z_{\sun}$) can be 
derived from (Bohlin et al. 1978): 
 \begin{equation} N({\rm H}_2) = 9.35\times 10^{20}\ A_{\rm V}\ Z^{-1} 
({\rm cm^{-2} mag^{-1}}) 
\end{equation} 
assuming a standard ratio of total to selective absorption of 3.1, that the 
column density of atomic hydrogen is negligible in the center of the galaxy, 
and that the amount of dust, and thus of extinction, increases with 
metallicity (Mauersberger et al. 1996, their Appendix B). The column density 
averaged over the entire extent of the bulge will be 2$N({\rm H}_2)$. 

\begin{table}
\caption[~\TNH2]{Column density of H$_2$ and CO--$N$(H$_2$) conversion factors
(in units of 10$^{20}$ cm$^{-2}$ (K km s$^{-1}$)$^{-1}$)}
{\small

\begin{flushleft}
\begin{tabular}{ccccccc}
\noalign{\smallskip}
\hline
\noalign{\smallskip}
NGC&$C_{{\rm H}\beta}$&$A_{\rm V}$&$N$(H$_2$)& $\alpha/\alpha_{\rm gal}$&
\multicolumn{2}{c}{$N$(H$_2$)/$I_{\rm CO}$}\\
&(mag)&(mag)& 10$^{20}$cm$^{-2}$ && obs.& stand.\\
\noalign{\smallskip}
\hline
3049& 0.32& 0.68& 10.6& 0.78& 1.00& 2.3\cr
5430& 1.63& 3.48& 50.1& 0.81& 1.00& 2.4\cr
6764& 0.87& 1.88& 43.5& 1.03& 1.73& 3.1\cr
\noalign{\smallskip}
\hline
\end{tabular}
\end{flushleft}

}
\end{table} 

$A_{\rm V}$, the total extinction in V (in magnitudes), is computed from 
$C_{{\rm H}\beta}$ and the following relation (Miller \& Mathews 1972): 
 \begin{equation} 
A_{\rm V} = 2.5\,C_{{\rm H}\beta}\,{\delta m_{\rm V} \over 
\delta m_{\beta}} = 2.136\,C_{{\rm H}\beta}.
\end{equation} 
Finally, $C_{{\rm H}\beta}$ and (O/H) (as metallicity indicator) are given in 
Table~\TOPTLINE\ for many star formation regions of the three Wolf-Rayet 
galaxies. Mean values over the central region are adopted. 

The results are presented in Table~\TNH2, where $N$(H$_2$) is derived from
the total extinction in V, $\alpha/\alpha_{\rm gal}$ is 
the correction factor for metallicity (Wilson 1995, Arimoto et al. 1996) and 
$N$(H$_2$)/$I_{\rm CO}$ is the conversion factor. The ``observed" value 
of the conversion factor was derived by the extinction method, from
$N$(H$_2$) and $I_{\rm CO}$ estimated in each galaxy, assuming a standard 
intensity ratio CO(2$\rightarrow$1)/CO(1$\rightarrow$0) of 0.87 (Braine \& 
Combes 1992) for optically thick gas in NGC 3049, and using the 
CO(1$\rightarrow$0) intensities given in Sect. 5.2 for the two other galaxies. 
The ``standard" value (predicted by the $\gamma$-ray method and assuming 
virial equilibrium) is the one of Strong et al. (1988), multiplied by 
$\alpha/\alpha_{\rm gal}$ to correct for metallicity effects in the galaxy 
(Wilson 1995, Arimoto et al. 1996).  The conversion factor predicted by the 
extinction method is between 2 and 3 times smaller than that predicted by the 
$\gamma$-ray method.  

\section{Discussion}

We have searched for dense molecular gas in three galaxies, NGC 3049, 5430 and  
6764, which share the following common properties.  They are the sites of very  
young ($\simeq$~5 Myr) starbursts, as evidenced by the presence of Wolf-Rayet  
stars and by our age determinations, they have a large far-infrared flux, they 
are barred and highly inclined. 

HCN has been detected in NGC~5430 and 6764, CS in the latter galaxy only. 
The HCN/FIR luminosity ratios of the two galaxies appear to 
be normal, confirming the trend noticed by Helfer \& Blitz (1993) for normal 
and starburst galaxies.  The upper limit to that ratio for NGC 3049 is also 
consistent with that trend.  The measured intensities of CS or upper limits 
are also normal relative to that of CO and HCN in the same galaxies, when 
compared to other surveys of CS in external galaxies.  If large amounts of 
dense molecular gas are required for star formation in burst mode, they no 
longer exist 5 Myr after the burst has started, presumably because they have 
been used up to form stars and/or ionized by the intense radiation emanating 
from massive hot stars. 

We also find that the HCN/CO integrated intensity ratios are rather low 
relative to the mean value found in other surveys of HCN in external 
galaxies. This is due to the fact that our three galaxies have an unusually 
high CO integrated intensity relative to their FIR flux, compared to other
galaxies, {\it including ones with young starbursts}.  In other words, this
higher than usual CO intensity is not a general property of galaxies
with young starbursts.  

The standard CO--$N$(H$_2$) conversion factor overestimates the amount of 
molecular hydrogen by a factor 2 or 3 in our three Wolf-Rayet galaxies.  
But again, our galaxies have unusual CO intensities.  This standard factor is 
thus probably valid for starburst galaxies in a statistical way, with a large 
uncertainty for individual estimates. 

Can the optical properties of our galaxies provide an explanation for their 
molecular line properties?  The {\it central} star formation rates in the 
three galaxies, estimated by the luminosity of the H$\alpha$ line, are not 
unusual among starburst galaxies, because they are all proportional 
(with the right factor) to the {\it global} star formation rates estimated by 
the FIR luminosities. The FIR luminosity and that of H$\alpha$ in the center 
are highest in NGC 5430 and lowest in NGC 3049, reflecting the pecking order 
for the luminosities in the molecular lines.  This is another confirmation 
that the latter are not unusual in very young starbursts. 

We note that NGC 5430, the galaxy with the oldest starburst, also has a 
circum-nuclear ring.  Its presence is consistent with the fact that most of 
the bar is in differential rotation (Sect. 4.1), as such rings form where the 
rotation becomes differential (Lesch et al. 1990).  The timescale of ring 
formation (a few 10$^8$ yr) is much larger than the age of the central 
starburst (6.0 Myr) of NGC 5430, and of the young star clusters (some less 
than 10 Myr) recently detected in other circum-nuclear rings by the {\it 
Hubble Space Telescope} (Maoz et al. 1996).  The absence of such a structure 
in NGC 3049 and 6764 should thus not be attributed to the fact that their 
central starbursts are younger, but to dynamical properties, as their bars 
appear to be rigidly rotating. 

One interesting property of NGC 5430 revealed in the present paper which 
deserves to be explored further is the fact that the CO and HI velocity 
profiles are asymmetric in opposite ways; there is relatively less HI where 
there is more CO.  One possible reason for this asymmetry is a more efficient 
conversion of HI to H$_2$ in the Wolf-Rayet region which may explain the 
existence of the young starburst at that end of the bar.  A detailed 
comparison of the relative distributions of HI and CO in starburst galaxies 
would certainly lead to a better understanding of the transformation of the 
gas before and during starbursts. 

Finally, we point out that the linewidths of the CO lines are correlated to 
the central starburst ages of the three galaxies.  This is a general property 
of young starbursts, which has been discovered by us (Contini et al. 1996a). 

\begin{acknowledgements}
Data from the literature were obtained with the Lyon Meudon Extragalactic 
database (LEDA), supplied by the LEDA team at CRAL-Observatoire de Lyon 
(France). We thank Bertrand Lefloch for assistance at the radiotelescope, and 
Rapha\"el Moreno for helpful comments on the paper. Remarks from an anonymous 
referee helped us provide a more quantitative discussion of our molecular line 
observations.  We also thank the staff of Observatoire du Pic du Midi and of 
Observatoire de Haute-Provence for assistance at the telescope.  H.W. thanks 
the French Academy of Sciences for financial support. 
\end{acknowledgements}

\end{document}